# Terahertz time-domain signatures of the inverse Edelstein effect in topological-insulator|ferromagnet heterostructures

G. Bierhance[1,2], C. In[1], E. Rongione[3,4], R. Rouzegar[1], O. Gueckstock[1], E. Longo[5,6], L. Baringthon[3], N. Reyren[3], R. Lebrun[3], J.-M. George[3], P. Tsipas[7], M. Wolf[2], T. S. Seifert[1], R. Mantovan[5], H. Jaffrès[3], A. Dimoulas[7], T. Kampfrath[1,2]

[1]Department of Physics, Freie Universität Berlin, 14195 Berlin, Germany
[2]Department of Physical Chemistry, Fritz Haber Institute of the Max Planck Society, 14195 Berlin, Germany
[3]Laboratoire Albert Fert, CNRS, Thales, Université Paris-Saclay, Palaiseau, F-91767 France
[4]Catalan Institute of Nanoscience and Nanotechnology (ICN2), CSIC and BIST, Campus UAB, Bellaterra, 08193 Barcelona, Spain
[5]Institute for Microelectronics and Microsystems, CNR-IMM Unit of Agrate Brianza, Via C. Olivetti 2, 20864 Agrate Brianza, Italy
[6]Institut de Ciència de Materials de Barcelona (ICMAB-CSIC), Campus de la UAB, 08193 Bellaterra, Spain
[7]Institute of Nanoscience and Nanotechnology, National Center for Scientific Research "Demokritos", Athens, G-15310 Greece

**Abstract**

Three-dimensional topological insulators possess topologically protected surface states with spin-momentum locking, which enable spin-charge-current interconversion (SCI) by the inverse Edelstein effect (IEE). However, it remains experimentally challenging to separate the surface-related IEE from the bulk-type inverse spin Hall effect (ISHE). Here, we search for distinct time-domain signatures of the two SCI phenomena in a $\mathcal{F}|\text{TI}$ model stack of a ferromagnetic-metal layer $\mathcal{F}$ (Co and Fe) and a topological-insulator layer TI ($Bi_2Te_3$, $SnBi_2Te_4$ and $Bi_{1-x}Sb_x$ with $x = 0.15$ and $0.3$), where the focus is on $Bi_2Te_3$. A femtosecond laser pulse serves to induce a transient spin voltage $\mu_s^\mathcal{F}$ in $\mathcal{F}$ and, thus, drive an ultrafast spin current out of $\mathcal{F}$. SCI results in a transverse charge current with a sheet density $I_c$ that is detected by sampling the emitted terahertz electric field. Analysis of the dynamics of $I_c(t)$ vs time $t$ relative to $\mu_s^\mathcal{F}(t)$ reveals two components with distinct time scales: (i) a quasi-instantaneous response and (ii) a longer-lived response with a relaxation time of 270 fs, which is independent of the chosen $\mathcal{F}$ material. Component (i) is consistently ascribed to the ISHE. In contrast, we interpret component (ii) as a signature of interfacial spin accumulation and the IEE at the $\mathcal{F}/Bi_2Te_3$ interface, with a fraction of $< 10^{-2}$ of the incident spins participating. This assignment is fully consistent with respect to its dynamics and magnitude. We rate other possible signal contributions, such as spin trapping in intermediate states, as less likely. Our results show that the femtosecond dynamics of photocurrents provide important insights into the mechanisms of spin transport and SCI in $\mathcal{F}|\text{TI}$ stacks.

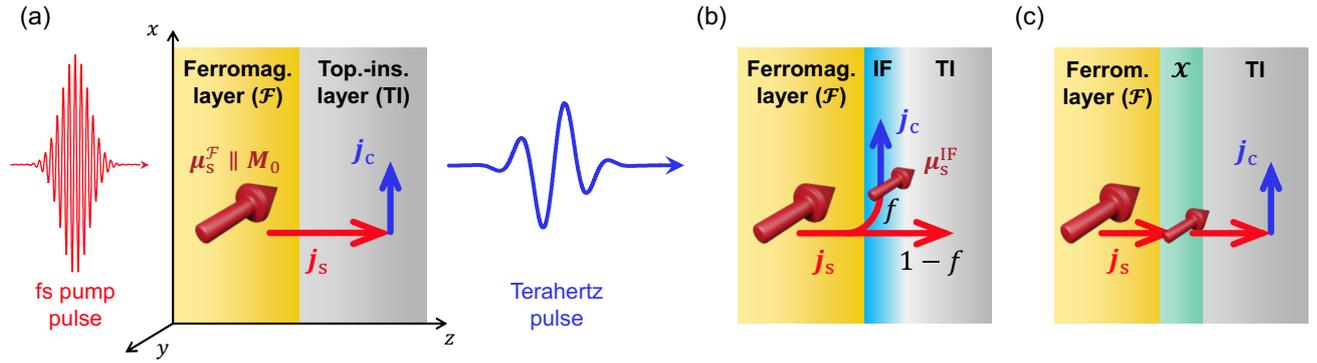

**Figure 1.** Schematic of mechanisms of spin transport and spin-charge-current interconversion in $\mathcal{F}|\mathrm{TI}$ stacks. **(a)** A spin current with density $j_\mathrm{s}$ flows from the ferromagnetic-metal layer $\mathcal{F}$ with magnetization $\boldsymbol{M}_0$ into a topological-insulator layer TI. The $j_\mathrm{s}$ is converted into a transverse charge current with density $j_\mathrm{c}$ by the inverse spin Hall effect (ISHE) in the TI bulk and/or **(b)** the inverse Edelstein effect (IEE) at the $\mathcal{F}$/TI interface (IF) that hosts a transient spin accumulation $\mu_\mathrm{s}^{\mathrm{IF}}$. Here, $f$ is the fraction of the incident $j_\mathrm{s}$ that is injected into the interfacial states and, thus, induces $\mu_\mathrm{s}^{\mathrm{IF}}$. **(c)** Before the fraction $1-f$ of $j_\mathrm{s}$ reaches the TI bulk, some of these spins may be trapped in an interlayer $\mathcal{X}$. As indicated in panel (a), the spin current is, in our experiment, driven by a femtosecond laser pulse through, e.g., the spin voltage $\mu_\mathrm{s}^{\mathcal{F}}$ that arises from the optical heating of electrons in $\mathcal{F}$. The $z$-integrated $j_\mathrm{c}$ is detected by measuring the emitted THz electromagnetic pulse.

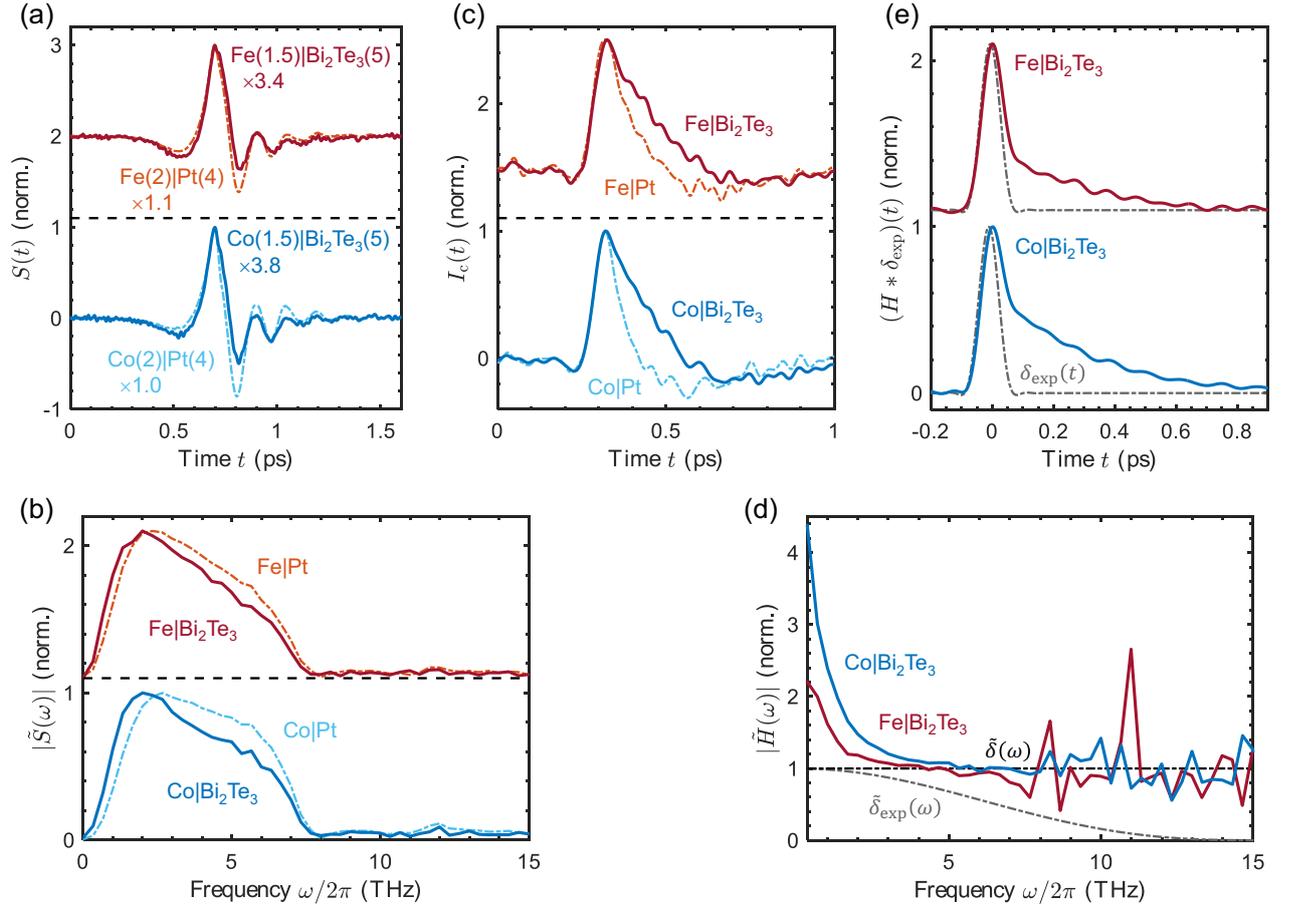

**Figure 2.** From THz signals $S$ to currents $I_c$ to response functions $H$ of $\mathcal{F}|Bi_2Te_3$ stacks. **(a)** Normalized electro-optic signals $S(t)$ of THz pulses emitted from the photoexcited samples. Solid lines show signals from $\mathcal{F}|Bi_2Te_3$, while dash-dotted lines indicate signals from the $\mathcal{F}|Pt$ reference, where $\mathcal{F}$ is Co or Fe. Layer thicknesses in nanometers are given by numbers in parentheses. All data is normalized to Co|Pt and scaled by the indicated factors. **(b)** Normalized magnitude of the Fourier amplitude $\tilde{S}(\omega)$ vs frequency $\omega/2\pi$ for the signals shown in panel (a). **(c)** Normalized charge-current sheet density $I_c(t)$ as extracted from the signals in panel (a) through Eq. (3). The $I_c(t)$ of $\mathcal{F}|Pt$ has the same temporal dynamics as the transient spin voltage $\mu_s^{\mathcal{F}}(t)$ of $\mathcal{F}$ = Fe and Co. **(d)** Fourier magnitude $|\tilde{H}(\omega)|$ of the $\mu_s^{\mathcal{F}}$-to-$I_c$ response function $H$ [see Eqs. (5) and (6)], normalized to an average of 1 over the interval 9-15 THz. The black dash-dotted line is the Fourier amplitude $\tilde{\delta}(\omega) = 1$ of the Dirac $\delta$-function, whereas $\tilde{\delta}_{\text{exp}}(\omega)$ (grey dash-dotted line) accounts for the finite bandwidth of our setup. **(e)** Normalized $\mu_s^{\mathcal{F}}$-to-$I_c$ response function $(H * \delta_{\text{exp}})(t)$ in the time domain. It summarizes the spin-transport and SCI dynamics following an impulsive spin voltage $\mu_s^{\mathcal{F}}(t) \propto \delta_{\text{exp}}(t)$ (grey dash-dotted line), which is the inverse Fourier transformation of $\tilde{\delta}_{\text{exp}}(\omega)$ in panel (d). In some panels, waveforms are vertically offset for clarity and horizontally shifted such that signal maxima coincide.

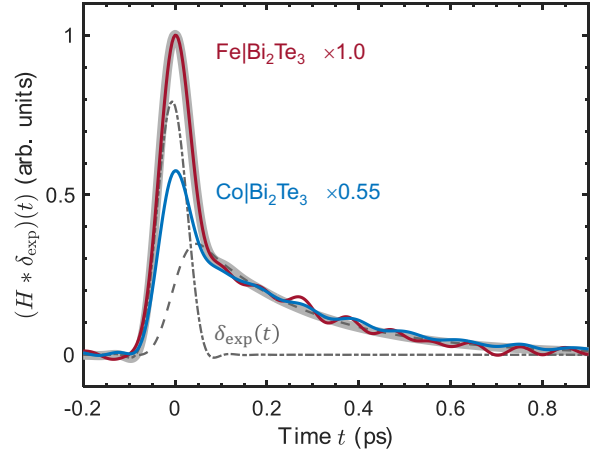

**Figure 3.** Fit to the $\mu_s^{\mathcal{F}}$-to-$I_c$ response functions $(H * \delta_{\mathrm{exp}})(t)$ of $\mathcal{F}|\mathrm{Bi}_2\mathrm{Te}_3$ with $\mathcal{F} = \mathrm{Fe}$ and Co [see Fig. 2(e)]. The modeled $H$ [Eq. (7)] for Fe|Bi$_2$Te$_3$ (thick grey line) contains the sum of a scaled instantaneous response $a\delta(t)$ (grey dash-dotted line) and a scaled single-sided exponential decay $bD_\tau(t)$ (grey dashed line). Both components include a convolution with the experimental $\delta$-function $\delta_{\mathrm{exp}}(t)$ [see Fig. 2(e)] that accounts for the finite bandwidth of our THz emission setup. The measured response functions are horizontally shifted such that signal maxima coincide. Note the scaling factors.

**I. Introduction**

Spin transport and spin-charge-current interconversion (SCI) are important functionalities in future spintronic devices. To keep pace with the speed of other information carriers, such as photons and electrons [1], one needs to push the speed of these functionalities to terahertz (THz) frequencies. An important spintronic model system is an $\mathcal{F}|\mathcal{N}$ stack consisting of a ferromagnetic metal layer $\mathcal{F}$ (e.g., Fe, Co, Ni) and a non-ferromagnetic layer $\mathcal{N}$ (e.g., Pt) [2, 3].

Two major SCI mechanisms in $\mathcal{F}|\mathcal{N}$ stacks are the inverse spin-Hall effect (ISHE) and the inverse Edelstein effect (IEE). The ISHE occurs in the $\mathcal{N}$ and/or $\mathcal{F}$ bulk regions [Fig. 1(a)] and converts a longitudinal spin current of density $j_\text{s}$ into a transverse charge current of density $j_\text{c}$ proportional to $j_\text{s}$, i.e., $j_\text{c} \propto j_\text{s}$. For the IEE, in contrast, a charge current arises from a transient spin accumulation (excess magnetization) $\mu_\text{s}^\text{IF}$ at an interface, e.g., between $\mathcal{F}$ and $\mathcal{N}$ [Fig. 1(b)], where the local inversion symmetry is broken. The IEE charge current is proportional to $\mu_\text{s}^\text{IF}$, i.e., $j_\text{c} \propto \mu_\text{s}^\text{IF}$.

In three-dimensional topological insulators (TIs), such as $Bi_2Te_3$ [4], noteworthy contributions to SCI are expected from both the ISHE [5] and IEE [6, 7]. In particular, topologically protected surface states (TSS) and spin-momentum locking at the $\mathcal{F}$/TI interface may lead to an enhanced IEE and, thus, a potential improvement of SCI efficiencies [8, 9]. At the same time, calculations predict that the contact between a ferromagnet $\mathcal{F}$, such as Co, and a TI, such as $Bi_2Se_3$, shifts the TSS below the Fermi level, thereby potentially distorting the helical spin texture [10-12]. Note that Rashba-type states of two-dimensional electron gases can also contribute to SCI at interfaces. In the following, we use the term IEE for SCI by both TSS and topologically trivial Rashba states.

As the $\mathcal{F}$/TI interface is buried, its electronic structure and, thus, possible IEE features remain hidden to electron probes like photoelectron emission spectroscopy [13]. Likewise, as the ISHE and IEE have identical macroscopic symmetry properties, they are hard to distinguish in transport experiments without measures like stacking modification [14]. However, ISHE and IEE may exhibit different temporal dynamics, which would, in principle, allow one to separate the two phenomena. For example, in heavy metals, the ISHE is expected to be instantaneous on a time scale of 50 fs and below because of the ultrashort relaxation time of electronic spin [15] and charge currents [16]. In contrast, the IEE has potentially slower dynamics because it requires an accumulation $\mu_\text{s}^\text{IF}$ of electron spins at the $\mathcal{F}/\mathcal{N}$ interface [Fig. 1(b)] and, thus, a temporal integration of the feeding spin current $j_\text{s}$. Consequently, we need to study SCI on the femtosecond scale to look for possible temporal signatures of the IEE. Ultrafast studies on $\mathcal{F}|$TI stacks based on THz emission spectroscopy were already conducted [6, 7, 17, 18], but these works focused on signal amplitudes rather than dynamics.

In this work, we use femtosecond laser pulses to drive spin transport in $\mathcal{F}|$TI stacks and $\mathcal{F}|$Pt reference stacks, where we choose Co and Fe for $\mathcal{F}$. The resulting photocurrent $j_\text{c}$ is measured by detecting the concomitantly emitted THz electromagnetic pulse [Fig. 1(a)]. We find that the photocurrent dynamics in $\mathcal{F}|$TI contain an additional slower component relative to $\mathcal{F}|$Pt. By assuming that $j_\text{c}$ is predominantly caused by the optically driven transient spin voltage of $\mathcal{F}$, we infer a response function that summarizes the temporal response of spin transport and SCI following a fictitious $\delta(t)$-like spin voltage. A non-instantaneous exponentially decaying response component with a time constant of 270 fs is found, which is independent of the choice (Co or Fe) of the $\mathcal{F}$ material. We consistently ascribe this contribution to non-instantaneous SCI by the IEE. The fraction of the spin current that is injected into the TSS is determined to be smaller than $10^{-2}$.

**II. Experimental setup**

To study spin transport and SCI on ultrafast time scales, we make use of broadband THz emission spectroscopy, as schematically shown in Fig. 1(a). In brief, a femtosecond laser pulse excites the sample, thereby driving an ultrafast longitudinal spin and, thus, transverse charge current that is probed by measuring the THz electromagnetic pulse it emits.

**A. Samples**

Our samples, all grown on Si substrates, are $\mathcal{F}|\mathcal{N}$ or $\mathcal{N}|\mathcal{F}$ stacks with $\mathcal{F} =$ Co or Fe. For $\mathcal{N}$, we choose the TIs $Bi_2Te_3$, $SnBi_2Te_4$ and $Bi_{1-x}Sb_x$ ($x = 0.15$ and $0.3$), and, as reference SCI material, Pt. Our main emphasis lies on the investigation of $Bi_2Te_3$. Note that the structures containing $Bi_2Te_3$ or $SnBi_2Te_4$ are grown on top of an InAs seed layer (thickness 30 nm) on the Si substrate (see Supplemental Material). Samples are characterized by DC conductivity measurements using a four-point probe system. More in-depth characterization of the same samples, e.g., by X-ray diffraction, angle-resolved photoelectron emission spectroscopy (ARPES) and scanning transmission electron microscopy, can be found in Refs. [17, 18].

**B. THz emission spectroscopy**

In the THz emission experiment [Fig. 1(a)], an external magnetic field >30 mT sets the direction of the static in-plane magnetization $\boldsymbol{M}_0$ of $\mathcal{F}$. An optical pump pulse (nominal duration 10 fs, center wavelength 800 nm, incident fluence up to 0.3 mJ/cm², repetition rate 80 MHz) from a Ti:sapphire laser oscillator transiently heats the metallic heterostructure and induces a transient excess of spin angular momentum in $\mathcal{F}$ [15]. This so-called spin voltage $\mu_s^{\mathcal{F}}$ and, potentially, other forces like temperature gradients [15] drive an ultrafast spin current with density $j_s$ from $\mathcal{F}$ to $\mathcal{N}$. SCI due, e.g., to the ISHE and IEE results in a transverse charge current $j_c$, which acts as a source of an electromagnetic pulse with frequencies extending into the THz range [19].

We measure the emitted THz electric field by electro-optic sampling with a linearly polarized gate pulse from the same laser in a GaP(110) crystal (thickness 250 µm, see Supplemental Material) [20, 21]. The resulting electro-optic signal $S(t, \boldsymbol{M}_0)$ is the THz-field-induced gate-pulse ellipticity as a function of the time delay $t$ between THz and gate pulse. We acquire THz signals for opposite $\mathcal{F}$ magnetizations $\pm \boldsymbol{M}_0$. Because we are interested in effects linear in the spin voltage $\mu_s^{\mathcal{F}}$, we focus on the signal component

$$S(t) = \frac{S(t, +\boldsymbol{M}_0) - S(t, -\boldsymbol{M}_0)}{2} \quad (1)$$

odd in $\boldsymbol{M}_0$. Signals even in $\boldsymbol{M}_0$ are typically one order of magnitude smaller than the odd-in-$\boldsymbol{M}_0$ signals for $\mathcal{F}|\mathcal{N}$ or $\mathcal{N}|\mathcal{F}$ stacks (Fig. S1). We observe no odd-in-$\boldsymbol{M}_0$ signal for a sample with $\mathcal{N} = Bi_2Te_3$ but without $\mathcal{F}$ (Fig. S2). Likewise, we confirm that the presence of an InAs seed layer between substrate and Fe has no impact on the signal dynamics obtained from the Fe|Pt reference sample (see Fig. S2).

**C. THz-photocurrent extraction**

The signal $S(t)$ contains electric-dipole and magnetic-dipole contributions [15]. The electric-dipole component arises from the charge-current with a sheet density

$$I_c(t) = \int dz\, j_c(z, t), \quad (2)$$

i.e., the charge-current density integrated along the stack normal [15, 22]. A nonzero $I_c$ requires broken inversion symmetry and includes the charge current due to SCI in the TI layer [Fig. 1(a)]. The magnetic-dipole contribution is induced predominantly by the optically induced magnetization dynamics in the ferromagnetic layer $\mathcal{F}$ [15, 23-25]. It is neglected here because it is typically one order of magnitude smaller than the signals measured.

With this assumption, the THz signal in the frequency domain is given by the product [15]

$$\tilde{S}(\omega) = \tilde{H}_{SE}(\omega)\tilde{Z}(\omega)\tilde{I}_c(\omega). \tag{3}$$

Here, the tilde denotes Fourier transformation of the relevant quantity, and $\omega/2\pi$ is frequency. The sample's sheet impedance $\tilde{Z}$ quantifies the conversion of $\tilde{I}_c$ into the THz electric field $\tilde{E} = \tilde{Z}\tilde{I}_c$ directly behind the sample. Finally, $\tilde{H}_{SE}$ captures the propagation of $\tilde{E}$ from the sample to the detector and the electro-optic detection process that result in the measured signal $\tilde{S}$.

By using Eq. (3), we can retrieve $\tilde{I}_c(\omega)$ and, thus, $I_c(t)$ [Eq. (2)] from the measured THz signal $\tilde{S}(\omega)$ [15, 22, 26]. To determine $\tilde{H}_{SE}(\omega)$, we choose the spintronic stack [27, 28] W(2 nm)|CoFeB(1.8 nm)|Pt(2 nm) as a reference emitter, whose current evolution is known [15, 29]. From THz-time-domain reflection spectroscopy, we conclude that the sheet impedance $\tilde{Z}(\omega)$ has a negligible frequency dependence (Fig. S3 and Supplemental Material). Therefore, we approximately determine $\tilde{Z}(\omega)$ from the DC sheet conductance (see Table S1).

### III. Results and discussion

#### A. THz raw signals

Figure 2(a) shows normalized THz signals $S(t)$ odd in $\boldsymbol{M}_0$ from $\mathcal{F}$(1.5 nm)|Bi$_2$Te$_3$(5 nm) and the $\mathcal{F}$(2 nm)|Pt(4 nm) reference stack, where $\mathcal{F}$ is made of Co or Fe. For Co(1.5 nm)|Bi$_2$Te$_3$(5 nm), we confirm explicitly that the THz signal depends linearly on the pump power (Fig. S4). Note that we multiply the signals from the TI samples with $-1$ to account for the reverse growth order of $\mathcal{F}$|Pt and Bi$_2$Te$_3$|$\mathcal{F}$.

While all waveforms in Fig. 2(a) have a somewhat similar shape, there are still notable qualitative differences. For example, the Co|Pt signal has sharper features than all other signals. This perception is reinforced by the magnitude of the Fourier-transformed signal $\tilde{S}(\omega)$ [Fig. 2(b)], which is enhanced at higher frequencies $\omega/2\pi$ within the accessible bandwidth compared to other signals. The spectral dip at $\omega/2\pi = 8$ THz is a characteristic feature of the electro-optic sampling in GaP [20].

#### B. THz charge currents

To enable a more direct comparison of the four THz signals in Fig. 2(a), we extract the charge-current evolution $I_c(t)$ from $S(t)$ (Section II.C) [15, 22]. As seen in Fig. 2(c), $I_c(t)$ rises instantaneously within our time resolution for all samples, but decays the fastest in Co|Pt, followed by Fe|Pt, Co|Bi$_2$Te$_3$ and, finally, Fe|Bi$_2$Te$_3$.

As shown previously [15], the photocurrent $I_c^{\mathrm{ref}}(t)$ in the Fe|Pt and Co|Pt reference samples monitors the dynamics of the spin voltage $\mu_s^{\mathcal{F}}(t)$ of $\mathcal{F}$, i.e.,

$$I_c^{\mathrm{ref}}(t) \propto \mu_s^{\mathcal{F}}(t). \tag{4}$$

Figure 2(c) reveals that the spin voltage rises step-like in both $\mathcal{F} = $ Co and Fe, but decays more rapidly in Co than in Fe. Assuming a biexponential-decay model [15], the dominant time constant, i.e., the electron-spin equilibration time, is 80 fs and 126 fs, respectively [30-33] (see Supplemental Material). This observation implies that spin relaxation in Co is substantially faster than in Fe [15].

#### C. Charge-current origin

Importantly, Fig. 2(c) indicates that the presence of $\mathcal{N} = $ Bi$_2$Te$_3$ slows down the dynamics of the photocurrent relative to $\mathcal{N} = $ Pt for both $\mathcal{F}$ choices, i.e., Fe and Co. To understand this slowing down, we recall that, for $\mathcal{N} = $ Pt, the THz emission signal was previously shown to arise from a spin current $j_s$ and its conversion into a charge current $j_c$ analogous to Fig. 1(a) [15]. As we observe THz emission for both Co|Bi$_2$Te$_3$ and Fe|Bi$_2$Te$_3$, we assume an analogous two-step process for the Bi$_2$Te$_3$ samples and briefly review SCI and spin transport in the following.

**SCI.** The charge-current density $j_c$ due to SCI generally scales linearly with the spin-current density $j_s(z,t)$ [Fig. 1(a)]. For metallic $\mathcal{N}$, SCI by the ISHE is quasi-instantaneous [16], resulting in $j_c(z,t) = \gamma_{\text{ISHE}}(z) j_s(z,t)$, where the spin-Hall angle $\gamma_{\text{ISHE}}$ quantifies the efficiency of this kind of SCI.

In contrast, for the IEE, SCI is localized at the $\mathcal{F}/\mathcal{N}$ interface. The resulting interfacial charge-current sheet density $I_c$ [Eq. (2)] is proportional to the instantaneous interfacial spin accumulation $\mu_s^{\text{IF}}$ through $I_c(t) \propto \mu_s^{\text{IF}}(t)$. Because a nonzero $\mu_s^{\text{IF}}$ requires an accumulation of the incident spin, we expect a time-integrating behavior of the feeding spin-current density $j_s$. In other words, $I_c(t)$ is expected to decay more slowly than $j_s(z,t)$, as also found in our measurements [Fig. 2(c)].

**Spin transport in $\mathcal{F}$|TI.** The spin current $j_s$ (Fig. 1) can originate from a number of processes. First, a direct photoinduced shift-type spin transfer at the $\mathcal{F}/\mathcal{N}$ interface [34, 35] would yield a very short derivative-like feature similar to $\partial \delta(t)/\partial t$ around the photocurrent onset [15], which is, however, not observed here [Fig. 2(c)]. We conclude that $j_s$ arises from band-like electron transport, which can, in general, be induced by gradients of electronic temperature (Seebeck-type contribution) or of spin voltage (pyrospintronic contribution) [15].

The Seebeck-type contribution is proportional to the difference between the electron temperature in $\mathcal{N}$ and the magnon or electron temperature in $\mathcal{F}$ [15, 22, 26, 36]. One can estimate the impact of the Seebeck-type contribution on $j_s$ by increasing the pump fluence. As the electron heat capacity of $\mathcal{N}$ = Bi$_2$Te$_3$ at 400 K is 30-40 times smaller than that of $\mathcal{F}$ = Fe and Co [15, 31, 37] (see Table S2), the electronic temperature in Bi$_2$Te$_3$ directly after sample excitation has a much more sublinear dependence on the pump fluence than in $\mathcal{F}$ (Fig. S5 and Supplemental Material). Consequently, the resulting Seebeck-type signal should depend very nonlinearly on the pump fluence for the fluences available in our experiment [38]. More precisely, the THz signal amplitude vs pump fluence would deviate from the initially linear trend by more than 70% already for an incident fluence of 0.1 mJ/cm$^2$. In contrast, we observe a highly linear dependence of the THz signal over the full range of pump fluences (Fig. S4). Therefore, a temperature difference between $\mathcal{F}$ and $\mathcal{N}$ contributes negligibly to $j_s$.

Thus, by exclusion, we deduce that the spin current in $\mathcal{F}$|Bi$_2$Te$_3$ with $\mathcal{F}$ = Co or Fe is solely driven by the pump-induced spin voltage in $\mathcal{F}$. Further, because the measured THz signals scale linearly with the pump fluence (Fig. S4), the spin-current density $j_s$, all spin accumulations and the resulting charge-current density $j_c$ are connected to the spin voltage $\mu_s^{\mathcal{F}}$ by a linear relationship. In particular, the charge-current sheet density $I_c$ [Eq. (2)] is given by a linear response to the spin voltage $\mu_s^{\mathcal{F}}$ and, thus, described by the convolution

$$I_c(t) = \left(H * \mu_s^{\mathcal{F}}\right)(t) = \int dt'\, H(t-t') \mu_s^{\mathcal{F}}(t'). \tag{5}$$

**Spin-transport/SCI response function.** In Eq. (5), $H(t)$ can be understood as the charge current $I_c(t)$ that is induced by an impulsive spin voltage, i.e., $\mu_s^{\mathcal{F}}(t') = \delta(t')$, which has a much simpler temporal structure than the actual spin voltage [Fig. 2(c)] [29]. The impulse response function $H$ summarizes all information on the subsequent spin-transport and SCI dynamics. We, thus, term it spin-transport/SCI or $\mu_s^{\mathcal{F}}$-to-$I_c$ response function of the given sample.

Importantly, any deviation of $H(t)$ from a $\delta(t)$-like shape is a signature of non-instantaneous dynamics of spin transport and/or SCI [see Fig. 2(c)]. An example of a quasi-instantaneous process is spin transport into Pt, where the lifetime of $j_s$ and any accumulation is < 15 fs [39]. A non-instantaneous spin-current response is the trapping of spin-polarized electrons in a layer like MgO between $\mathcal{F}$ and $\mathcal{N}$ [29]. Because of its much simpler interpretation, we focus on the dynamics of $H$ rather than of signals $S$ [Fig. 2(a)] or currents $I_c$ [Fig. 2(c)] in the following.

### D. Analysis of response function $H$

**Frequency domain.** To extract the $\mu_s^\mathcal{F}$-to-$I_c$ response function $H$ of a $\mathcal{F}|\text{Bi}_2\text{Te}_3$ stack, we Fourier-transform Eq. (5) into the frequency domain and obtain the simple product $\tilde{I}_c(\omega) = \tilde{H}(\omega)\tilde{\mu}_s^\mathcal{F}(\omega)$. For the $\mathcal{F}|\text{Pt}$ reference sample, one has $I_c^{\text{ref}}(t) \propto \mu_s^\mathcal{F}(t)$ [15, 29], and, consequently, $\tilde{H}^{\text{ref}}(\omega)$ is constant over the whole bandwidth (0.3-15 THz) of our experiment. By applying Eq. (3) to both the sample signal $\tilde{S}(\omega)$ ($\mathcal{F}|\text{TI}$) and the reference signal $\tilde{S}^{\text{ref}}(\omega)$ ($\mathcal{F}|\text{Pt}$), we find

$$\tilde{H}(\omega) \propto \frac{\tilde{I}_c(\omega)}{\tilde{\mu}_s^\mathcal{F}(\omega)} \propto \frac{\tilde{S}(\omega)/\tilde{Z}(\omega)}{\tilde{S}^{\text{ref}}(\omega)/\tilde{Z}^{\text{ref}}(\omega)}, \tag{6}$$

where $\tilde{Z}^{\text{ref}}(\omega)$ is the THz sheet impedance of the $\mathcal{F}|\text{Pt}$ reference. Importantly, the instrument response function $\tilde{H}_{SE}(\omega)$ [Eq. (3)] cancels in Eq. (6). Further, $\tilde{Z}^{\text{ref}}(\omega)$ and $\tilde{Z}(\omega)$ are independent of frequency to very good approximation (Fig. S3 and Supplemental Material). Therefore, $\tilde{H}(\omega)$ is, apart from a global scaling factor, obtained by straightforward division of THz emission signals from the $\mathcal{F}|\text{Bi}_2\text{Te}_3$ and $\mathcal{F}|\text{Pt}$ samples in the frequency domain [Eq. (6)], without the need to measure any instrument response functions.

Note that $\tilde{S}(\omega)$ and $\tilde{S}^{\text{ref}}(\omega)$ exhibit a dip at 8 THz [Fig. 2(b)], which arises from a zero of the electro-optic response of our GaP detector [20]. To avoid division by the very small and, thus, relatively uncertain $\tilde{S}^{\text{ref}}(\omega)$ in Eq. (6), we determine $\tilde{H}(\omega)$ at 8 THz by interpolation of $\tilde{H}(\omega)$ at the adjacent frequencies.

The modulus of the extracted $\mu_s^\mathcal{F}$-to-$I_c$ response function $\tilde{H}(\omega)$ of $\text{Co}|\text{Bi}_2\text{Te}_3$ and $\text{Fe}|\text{Bi}_2\text{Te}_3$ is displayed in Fig. 2(d). The magnitude of $\tilde{H}(\omega)$ decreases from 0.3 to 5 THz and subsequently remains constant up to the maximum covered frequency of 15 THz. Therefore, the range 5-15 THz appears to be dominated by an instantaneous, scaled $\delta(t)$-like response with constant Fourier transformation $\tilde{\delta}(\omega) = 1$ [Fig. 2(d)]. In contrast, in the interval 0.3-5 THz, $\tilde{H}(\omega)$ has significant variation, which suggests non-instantaneous contributions to the dynamics of spin current and/or SCI.

**Time domain.** It is highly instructive to consider the $\mu_s^\mathcal{F}$-to-$I_c$ response functions of $\text{Co}|\text{Bi}_2\text{Te}_3$ and $\text{Fe}|\text{Bi}_2\text{Te}_3$ in the time domain. Prior to the inverse Fourier transformation, we multiply $\tilde{H}(\omega)$ with a filter function $\tilde{\delta}_{\text{exp}}(\omega)$ to account for the finite usable bandwidth of our data [Fig. 2(d)]. We choose $\tilde{\delta}_{\text{exp}}(\omega)$ to be a Norton-Beer window function with a cutoff at $\omega/2\pi = \pm 15$ THz [40, 41] because its time-domain counterpart $\delta_{\text{exp}}(t)$ is almost unipolar with a full width at half maximum (FWHM) of 74 fs [Fig. 2(e)]. By inverse Fourier transformation of the filtered $\tilde{H}(\omega)\tilde{\delta}_{\text{exp}}(\omega)$, we obtain $(H * \delta_{\text{exp}})(t)$ in the time domain. The convolution of the response function $H(t)$ with the unipolar peak $\delta_{\text{exp}}(t)$ possibly smoothens features of $H$ that are finer than the width of $\delta_{\text{exp}}$, but otherwise leaves the shape of $H$ unchanged.

The extracted $(H * \delta_{\text{exp}})(t)$ of both $\text{Co}|\text{Bi}_2\text{Te}_3$ and $\text{Fe}|\text{Bi}_2\text{Te}_3$ is zero for $t < 0$ [Fig. 2(e)], as expected for a causal response. At time $t \approx 0$ ps, it turns into a peak with a shape similar to $\delta_{\text{exp}}(t)$, which characterizes the time resolution with which $H$ is determined. Importantly, the scaled relaxation tail of $H * \delta_{\text{exp}}$ is identical within the accuracy of our experiment for both $\text{Co}|\text{Bi}_2\text{Te}_3$ and $\text{Fe}|\text{Bi}_2\text{Te}_3$ (Fig. 3).

Consequently, we tentatively parameterize the measured response function by

$$H(t) = a\delta(t) + bD_\tau(t) \quad \text{with} \quad D_\tau(t) = \frac{1}{\tau}\Theta(t)e^{-t/\tau}. \tag{7}$$

It is the sum of a $\delta$-function and a single-sided exponential decay $D_\tau(t)$ with $\int \mathrm{d}t\, D_\tau(t) = 1$, where $\Theta(t)$ denotes the Heaviside step function. The weight coefficients $a$ and $b$ and the time constant $\tau$ are free parameters. Figure 3 shows that a fit using $H * \delta_{\exp}$ with $H$ from Eq. (7) excellently reproduces our experimental results for $\tau \approx 270 \pm 20$ fs for the $\mathcal{F}|$TI stacks with the TI Bi$_2$Te$_3$.

Let us summarize this significant result. Within our experimental accuracy, the charge-current response $H(t)$ induced by an impulsive spin voltage $\mu_s^{\mathcal{F}}(t) = \delta(t)$ in $\mathcal{F}$ is the sum of an instantaneous response $[a\delta(t)]$ and an exponential decay $[bD_\tau(t)]$. Remarkably, when the spin-current source $\mathcal{F} = $ Co is replaced by Fe, only the amplitudes $a$ and $b$ change, whereas the changes in the relaxation time $\tau$ are minor. This finding indicates that the photocurrent $I_c(t)$ arises from a superposition of two processes of spin transport and/or SCI in Bi$_2$Te$_3$ rather than in $\mathcal{F}$. The first process $[a\delta(t)]$ is instantaneous within our time resolution, whereas the second one $[aD_\tau(t)]$ has a relaxation time of $\tau \approx 270$ fs.

Fig. S6 shows the corresponding response function $(H * \delta_{\exp})(t)$ for Co|SnBi$_2$Te$_4$ and Co|Bi$_{1-x}$Sb$_x$ samples with varying thickness and composition ($x = 0.15$ and $0.3$). SnBi$_2$Te$_4$ displays very similar dynamics in comparison to Bi$_2$Te$_3$, in line with the model of Eq. (7). In contrast, Bi$_{1-x}$Sb$_x$(5 nm) shows a faster and more complex response (see Supplemental Material).

**E. Interpretation: ISHE and IEE**

Based on the previous analysis and discussion, we suggest the following microscopic scenario to explain our observations. First, if we assume a $\delta$-like spin voltage, it drives a $\delta$-like spin current across the $\mathcal{F}/$TI interface [Fig. 1(b)]. A fraction $f$ of this current ($0 \leq f \leq 1$) is coupled into the interfacial TSS and/or Rashba-type states, whereas the remaining part $1 - f$ flows into the TI bulk [Fig. 1(b)].

We tentatively assign $H_{\mathrm{ISHE}} \coloneqq a\delta$ in Eq. (7) to the ISHE in the bulk of $\mathcal{F}$ and Bi$_2$Te$_3$, whereas $H_{\mathrm{IEE}} \coloneqq bD_\tau$ is ascribed to the IEE at the $\mathcal{F}/$Bi$_2$Te$_3$ interface, which is slower because of the required spin accumulation $\mu_s^{\mathrm{IF}}$ [Fig. 1(b)]. The coefficients $a$ and $b$ are proportional to $1 - f$ and $f$, respectively. Importantly, this interpretation is consistent with respect to the sign, magnitude and temporal shape of ISHE and IEE, as discussed in the following.

**ISHE.** The $\delta$-like spin voltage in $\mathcal{F}$ injects spin-polarized electrons into the TI bulk, where they propagate with the Fermi velocity $v_{\mathrm{F}}^{\mathrm{B}}$ into the depth of the film. This ballistic current relaxes on the time scale of the bulk momentum-relaxation time $\tau^{\mathrm{B}}$ and is, simultaneously, converted into an in-plane charge current by the ISHE. Therefore, $H_{\mathrm{ISHE}}$ is proportional to the product $(1 - f)v_{\mathrm{F}}^{\mathrm{B}}\theta_{\mathrm{SH}}$, where $\theta_{\mathrm{SH}}$ is the spin Hall angle of the TI and relaxes according to $\Theta(t)e^{-t/\tau^{\mathrm{B}}} \approx \tau^{\mathrm{B}}\delta(t)$. The latter approximation is justified because the bulk Drude scattering time $\tau^{\mathrm{B}}$ [16] is significantly shorter than our time resolution of 74 fs. Examples are ferromagnets like Fe ($\tau^{\mathrm{B}} = 30$ fs) [42, 43] and typical TIs like Bi$_2$Te$_3$ ($\tau^{\mathrm{B}} < 40$ fs) [44]. We, therefore, obtain

$$H_{\mathrm{ISHE}}(t) \propto (1 - f)v_{\mathrm{F}}^{\mathrm{B}}\tau^{\mathrm{B}}\theta_{\mathrm{SH}}\delta(t) \tag{8}$$

in the ballistic limit, which is relevant on ultrafast time scales [45]. We note that Eq. (8) is fully consistent with the $a\delta$ term in Eq. (7). Further, the sign of $a$ and, therefore, $\theta_{\mathrm{SH}}$ is positive, analogous to Pt [Fig. 2(c)]. This finding agrees with the sign of the spin Hall angle from previous works on Bi$_2$Te$_3$ grown by molecular-beam epitaxy [46]. We note that a possible ISHE in Fe and Co may contribute to $a$, but the ISHE in TIs dominates because the spin-Hall angle of Fe and Co is negative relative to Pt [47].

**IEE.** Here, the $\delta$-like spin voltage in $\mathcal{F}$ injects spin angular momentum into TSS or Rashba-type states of the $\mathcal{F}/$TI interface [Fig. 1(b)]. Due to spin-momentum locking of these states, each excess spin parallel to $\boldsymbol{M}_0$ is accompanied by one electron moving with the interface Fermi velocity $v_{\mathrm{F}}^{\mathrm{IF}}$ in the interface plane but perpendicular to $\boldsymbol{M}_0$ [10]. Therefore, the IEE-related charge current $H_{\mathrm{IEE}}$ of TSS scales with $f v_{\mathrm{F}}^{\mathrm{IF}}$.

To model the dynamics of $H_{\mathrm{IEE}}$, we note that the total electron spin in the $\mathcal{F}$/TI interface is proportional to the interface spin accumulation $\mu_s^{\mathrm{IF}}$. The latter decays with a rate $\partial \mu_s^{\mathrm{IF}}/\partial t = -\mu_s^{\mathrm{IF}}/\tau^{\mathrm{IF}}$ due, e.g., to spin-conserving transfer (tr) into $\mathcal{F}$ or TI bulk states or, possibly, spin-flip (SF) scattering with a total inverse time constant $(\tau^{\mathrm{IF}})^{-1} = \tau_{\mathrm{tr}}^{-1} + \tau_{\mathrm{SF}}^{-1}$, resulting in $\mu_s^{\mathrm{IF}}(t) \propto \Theta(t)\mathrm{e}^{-t/\tau^{\mathrm{IF}}}$. We, thus, find

$$H_{\mathrm{IEE}}(t) \propto f v_{\mathrm{F}}^{\mathrm{IF}} \tau^{\mathrm{IF}} D_{\tau^{\mathrm{IF}}}(t). \tag{9}$$

Eq. (9) is in line with the $b D_\tau$ term in Eq. (7). Moreover, the sign of $b$ and, thus, the assumed IEE is positive, analogous to the sign of the ISHE of Pt. This finding may suggest that $b D_\tau$ in Eq. (7) arises from the IEE of the TSS rather than trivial Rashba states [6]. The reason is that these two kinds of states usually contribute with opposite sign to SCI, as inferred from ARPES measurements on similar TIs like Bi$_2$Se$_3$ and from spin-transfer ferromagnetic-resonance measurements on (Bi$_{0.4}$Sb$_{0.6}$)$_2$Te$_3$ [48-50]. However, exceptions of this notion were already observed [51].

Remarkably, the characteristic time $\tau^{\mathrm{IF}}$ of 270 fs found in our experiment (Fig. 3) is compatible with interface spin relaxation times measured by time-resolved ARPES in Sb$_2$Te$_3$ samples under similar excitation conditions [52, 53]. Further, THz-transmission and THz-reflection spectroscopy studies of Bi$_2$Te$_3$ suggest scattering times around 200 fs for the surface states at room temperature [44, 54]. Note, however, that these time constants were determined for TI samples without any adjacent layer like $\mathcal{F}$.

**Estimation of $f$.** It is instructive to estimate the fraction $f$ of the spin current from $\mathcal{F}$ to TI that is coupled into the TSS [Fig. 1(b)]. To this end, we note that, in the measured $H * \delta_{\mathrm{exp}}$, the area under the peak ($a\delta$) and the relaxation tail ($b D_\tau$) is roughly equal (Fig. 3). This observation along with our interpretation implies $\int \mathrm{d}t\, H_{\mathrm{ISHE}}(t) \sim \int \mathrm{d}t\, H_{\mathrm{IEE}}(t)$. By using Eqs. (8) and (9), we conclude that

$$\frac{f}{1-f} \sim \frac{v_{\mathrm{F}}^{\mathrm{B}} \tau^{\mathrm{B}} \theta_{\mathrm{SH}}}{v_{\mathrm{F}}^{\mathrm{IF}} \tau^{\mathrm{IF}}}. \tag{10}$$

To estimate $f$, we take the reported bulk Drude scattering time $\tau^{\mathrm{B}} < 40$ fs of Bi$_2$Te$_3$ [44], while we use $\tau^{\mathrm{IF}} = 270$ fs as found in our experiment (Fig. 3). The Fermi velocity associated with the TSS is $v_{\mathrm{F}}^{\mathrm{IF}} = 0.4$ nm/fs [55, 56], and for the bulk Fermi velocity, we make the reasonable assumption $v_{\mathrm{F}}^{\mathrm{B}} < v_{\mathrm{F}}^{\mathrm{IF}}$. Together with the calculated spin Hall angle $\theta_{\mathrm{SH}} = 0.06$ of Bi$_2$Te$_3$ [57], Eq. (10) yields $f < 10^{-2}$.

Interestingly, the inferred $f \ll 1$ implies a relatively small probability that an electron spin traversing the interface from $\mathcal{F}$ to TI is transferred into the TSS. It, thus, suggests a relatively weak coupling of $\mathcal{F}$ and the TSS of the adjacent TI. This conclusion is consistent with the long lifetime $\tau^{\mathrm{IF}} = 270$ fs that is comparable to Bi$_2$Te$_3$ samples without any layer on top [44, 54].

**Static case.** As a final check, we extend our considerations to the quasi-static regime and calculate the SCI efficiency for the IEE as $\lambda_{\mathrm{IEE}} = \int \mathrm{d}t\, I_c / \int \mathrm{d}t\, j_s = \int \mathrm{d}t\, H_{\mathrm{IEE}}(t)$. We find $\lambda_{\mathrm{IEE}} = f v_{\mathrm{F}}^{\mathrm{IF}} \tau^{\mathrm{IF}}$, in agreement with the result $\lambda_{\mathrm{IEE}} = v_{\mathrm{F}}^{\mathrm{IF}} \tau'$ for static currents [58], if we identify $\tau' = f \tau^{\mathrm{IF}}$.

For $v_{\mathrm{F}}^{\mathrm{IF}} = 0.4$ nm/fs [55, 56], $f < 10^{-2}$ and $\tau^{\mathrm{IF}} = 270$ fs as inferred above, we obtain $\lambda_{\mathrm{IEE}} < 1$ nm. This value is consistent with the $\lambda_{\mathrm{IEE}} \sim 0.1$ nm [59-61] as obtained in experiments with spin pumping of $j_s$ and contact-based detection of the resulting static current $I_c$ for similar TIs [8, 9, 62].

In summary, our assignment of $a\delta$ in Eq. (7) to the ISHE in the bulk of $\mathcal{F}$ and Bi$_2$Te$_3$ [Fig. 1(a)] and of $b D_\tau$ to the IEE at the $\mathcal{F}$/Bi$_2$Te$_3$ interface [Fig. 1(b)] is highly consistent with respect to the amplitude and the shape of the dynamics of the measured response function $H * \delta_{\mathrm{exp}}$ (Fig. 3).

## F. Alternative contributions and mechanisms

Even though our interpretation of our experimental results is consistent, other mechanisms may also explain the observed current dynamics in Fig. 2(c). According to Eq. (5) and the remarks preceding it, a non-instantaneous modification $\propto D_\tau(t)$ [Eq. (7)] of the photocurrent relative to the reference $\mathcal{N} = $ Pt may not only arise from (1) changes in the SCI as discussed in Section III.E, but also from (2) changes in the spin-voltage dynamics and (3) the spin-current propagation.

Regarding scenario (2), the spin-voltage dynamics $\mu_s^\mathcal{F}(t)$ in $\mathcal{F}$ may change when $\mathcal{N} = $ Pt is replaced by $Bi_2Te_3$ due to a modified spin transfer from $\mathcal{F}$ to $\mathcal{N}$, thereby invalidating Eq. (6). This scenario is unlikely because spin transport into $\mathcal{N}$ was shown to only negligibly change the spin-voltage dynamics of $\mathcal{F}$ [15]. Alternatively, $\mu_s^\mathcal{F}(t)$ can be modified by a change in the electron-temperature dynamics in $\mathcal{F}$ due to heat flow from $\mathcal{N}$ to $\mathcal{F}$. In this case, the dynamics would sensitively depend on the difference of electronic temperature between $\mathcal{F}$ and $\mathcal{N} = Bi_2Te_3$. As discussed in Section III.C, a highly nonlinear fluence dependence of the THz signal would result from this hypothesis, which is not observed.

Concerning scenario (3), the non-instantaneous contribution $D_\tau$ may arise from a longer-lived ballistic spin-current propagation in $\mathcal{N} = Bi_2Te_3$. This assumption is, however, inconsistent with previous work [5] which found that $j_s(z,t)$ in $Bi_2Te_3$ relaxes over as little as 2 nm. Moreover, the characteristic bulk Drude scattering time of $Bi_2Te_3$ is $< 40$ fs and, thus, below $\tau = 270$ fs [44]. Alternatively, $D_\tau$ may result from the build-up of a longer-lived spin accumulation in the bulk of $\mathcal{N} = Bi_2Te_3$. In this scenario, a nonnegative $\delta$-like spin voltage in $\mathcal{F}$ would lead to a nonnegative spin voltage in $\mathcal{N}$. The latter, in turn, would drive spin back to $\mathcal{F}$ and imply a longer-lived spin-current component, but with opposite sign compared to $a\delta$, which is not observed.

A final option of scenario (3) would rely on structural imperfections of the $\mathcal{F}|Bi_2Te_3$ sample, i.e., a third layer $\mathcal{X}$ between $\mathcal{F} = $ Co or Fe and $\mathcal{N} = Bi_2Te_3$ [Fig. 1(c)]. The emerging localized states of $\mathcal{X}$ could partially trap spin-polarized electrons from $\mathcal{F}$ and later release them into $\mathcal{N}$. Indeed, such relaxation behavior on time scales of 100 fs was recently observed for MgO tunnel barriers between CoFeB and Pt thin films [29]. In our $Co|Bi_2Te_3$ sample, the existence of a third intermixing layer $\mathcal{X}$ is not implausible: Previous works observed the formation of a magnetically dead layer and interdiffusion between ferromagnets like Fe or Co and TIs like $Bi_2Se_3$ [63, 64] or $Sb_2Te_3$ [65, 66]. Likewise, for $Co|Bi_2Te_3$, the formation of interfacial $CoTe_2$ was proposed [67]. Similarly, at interfaces between ferromagnetic-metal and organic-molecule layers, spins can be trapped for up to 1 ps in hybrid interface states [68]. Possibly, such trapping could even occur in $\mathcal{F}$/TI interface states with negligible SCI. In any case, one would expect that the trap layer $\mathcal{X}$ and, thus, the trapping time depend sensitively on the adjacent layer $\mathcal{F}$. However, as almost identical time constant of $\tau \approx 270 \pm 20$ fs of the delayed feature $D_\tau$ is observed for both $\mathcal{F} = $ Co and Fe (Fig. 3), we consider scenario (3) as rather unlikely.

A short discussion of the $\mu_s^\mathcal{F}$-to-$I_c$ response functions of $Co|SnBi_2Te_4$ and $Co|Bi_{1-x}Sb_x$ can be found in the Supplemental Material.

## IV. Conclusions

In conclusion, we use THz-emission spectroscopy to investigate ultrafast spin-transport and SCI dynamics in $\mathcal{F}|Bi_2Te_3$ stacks. Extraction of a suitable response function permits a detailed quantitative analysis of these processes based on their temporal signatures. It reveals a non-instantaneous relaxation component with a relaxation time of 270 fs that is independent of the chosen $\mathcal{F}$ material Fe or Co. It can consistently be ascribed to spin accumulation and SCI by the IEE at the $\mathcal{F}$/TI interface [Fig. 1(b)]. In general, on the one hand, our study may serve as a blueprint to disentangle the complex spin transport and SCI responses in spintronic multilayers and, more generally, other conversion processes, such as trap and release [Fig. 1(c)]. On the other hand, this work reveals an approach toward emitters and detectors of broadband THz electromagnetic radiation [69, 70] whose response can be tailored by interface tuning.

**Acknowledgments**

We thank S. Valenzuela for fruitful discussions. We acknowledge funding by the European Union H2020 program through the FET project SKYTOP/Grant No. 824123, the ERC-2023 Advanced Grant ORBITERA (grant no. 101142285), and the German Research Foundation through the Collaborative Research Center SFB TRR 227 "Ultrafast spin dynamics" (project ID 328545488, projects A05 and B02).